\pdfoutput=1
\documentclass[10pt,oneside,pdftex,dvipsnames,a4paper]{article}

\usepackage{etex}
\reserveinserts{18}

\usepackage{a4wide}
\usepackage{setspace,xspace,makeidx,fancyvrb,relsize}
\usepackage{alltt}

\usepackage{amsmath,amsthm,amssymb}
\usepackage{parallel,amsfonts,accents}
\usepackage{amsxtra,amstext,latexsym,dsfont} % for \mathds{N}
\normalfont
\usepackage{bm}

\usepackage[comma]{natbib}
\usepackage[symbol]{footmisc}
\usepackage{perpage}
\MakePerPage[2]{footnote}

\usepackage[pdftex]{graphicx}
\usepackage[dvipsnames]{xcolor}

\usepackage{pgf,tikz}
\usetikzlibrary{snakes}
\usetikzlibrary{backgrounds}
\usetikzlibrary{arrows}
\usetikzlibrary{patterns}
\usetikzlibrary{fit}
\usetikzlibrary{calc}

\usepackage{lscape} 
\usepackage{shorttoc}

\usepackage{array,longtable,booktabs,float,rotating}
\usepackage{dcolumn,ctable}

\usepackage[labelfont={bf},ruled,labelsep=period,small]{caption}

\usepackage[flushleft]{threeparttable}
\usepackage[linkcolor=black,colorlinks=true,citecolor=black,%ForestGreen,%pagebackref,%
bookmarks=true,bookmarksopen=true,bookmarksnumbered=true,bookmarksopenlevel=0,%
hyperindex=true,pdfpagemode=UseOutlines,linktocpage=true,pdfpagelayout=TwoColumnLeft,
pdftitle=Biostatistics,pdfstartview=FitH,pdffitwindow=true,urlcolor=blue]{hyperref}
\usepackage{ragged2e}

\usepackage{multirow,multicol}

\usepackage{fancyhdr,extramarks}

 \makeatletter%--------------------------
 \newcommand\sub{\@startsection%
     {subsubsection}{5}{0mm}{-1\baselineskip}{.01\baselineskip}%
     {\normalfont\itshape}}
 \renewcommand\subsubsection{\@startsection%
     {subsubsection}{3}{0mm}{-1\baselineskip}{.01\baselineskip}%
     {\normalfont\itshape}}
 \makeatother%--------------------------

\makeatletter
        \newcommand\Appendix[2][?]{%
            \refstepcounter{section}%
            \addcontentsline{toc}{appendix}%
                {\protect\numberline{\appendixname~\thesection}#1}%
            {\raggedleft\bfseries \appendixname\
                \thesection\par \centering#2\par}%
                \sectionmark{#1}%
                \@afterheading
                \addvspace{\baselineskip}}
        \newcommand\sAppendix[1]{%
            \raggedleft\bfseries\appendixname\par
            \@afterheading\addvspace{\baselineskip}}
    \makeatother

\normalsize \makeindex

\newcolumntype{A}{>{\centering}p{100pt}}
\newlength\savedwidth

\def\coldot{.}%
{\catcode`\.=\active%
    \gdef.{$\egroup\setbox2=\hbox to \dimen0 \bgroup$\coldot}}
\def\rightdots#1{%
    \setbox0=\hbox{$1$}\dimen0=#1\wd0%
    \setbox0=\hbox{$\coldot$}\advance\dimen0 \wd0%
    \setbox2=\hbox to \dimen0 {}%
    \setbox0=\hbox\bgroup\mathcode`\.="8000 $}
\def\endrightdots{$\hfil\egroup\box0\box2}

\newcolumntype{d}[1]{D{.}{.}{#1}}
\newcolumntype{A}{>{\centering}p{100pt}}
\newcolumntype{.}{D{.}{.}{-1}}
\newcolumntype{P}[2]{>{#1\raggedright\arraybackslash}p{#2}}

\DeclareFontFamily{U}{euc}{}% I chose euc because the chart is called Euler cursive
\DeclareFontShape{U}{euc}{m}{n}{<-6>eurm5<6-8>eurm7<8->eurm10}{}%
\theoremstyle{plain}      
\theoremstyle{plain}      
\theoremstyle{plain}      
\theoremstyle{definition} 
\theoremstyle{definition} 
\theoremstyle{definition} 
\theoremstyle{plain} \newtheorem{cor}{Corollary}
\theoremstyle{definition} 
\theoremstyle{plain} \newtheorem{pro}{Proposition}
\theoremstyle{definition} 

\newcounter{nctr}
\usepackage[rightbars,color]{changebar}
\cbcolor{blue}
\VerbatimFootnotes
\usepackage{url}
\newcommand\tb{\textbf}
\newcommand\ti{\textit}

\newcommand\ds{\mathds}
\newcommand\bb{\mathbb}

\newcommand\te{\text}
\newcommand\ma[1]{\te{\bf{#1}}}
\newcommand\ca{\mathcal}

\newcommand\op{\operatorname}

\newcommand\argmin{\operatornamewithlimits{argmin}}

\newcommand\E{\bb{E}}

\newcommand\ind{\te{ind}}

\newcommand\lb{\lbrace}
\newcommand\lt{\left}

\newcommand\p{\bb{P}}           % Probability P
\newcommand\q{\quad}
\newcommand\qq{\qquad}

\newcommand\rb{\rbrace}
\newcommand\rt{\right}

\newcommand\stack{\stackrel} % Over equal signs.

\newcommand\tth{^\text{th}}

\newcommand\R{\ds{R}}  % Real
\newcommand\by{\ma{y}}

\newcommand\cI{\ca{I}} % Fisher Information
\newcommand\ga{\gamma}

\newcommand\bsig{\bm\sigma}

\newcommand\bth{\bm\theta}
\newcommand\bxi{\bm\xi}

\newcommand\bTh{\bm\Theta}

\makeatletter%--------------------------
\renewcommand\subsection{\@startsection%
    {subsection}{3}{0mm}{-1.5\baselineskip}{.1\baselineskip}%
    {\normalfont\large\itshape}}
\makeatother%--------------------------

\usepackage{fancyhdr}
\begin{document}
\sloppy

%-----------------------------------------------
% Title Page
\begin{center}
Running Head: \uppercase{Classification Loss Function} %25 characters
\end{center}
\vspace{3cm}

\begin{center}
\Large{\tb{Classification Loss Function for Parameter Ensembles \\ 
           in Bayesian Hierarchical Models.}} 
\\
%\vspace{2.5cm}\normalsize Authors et al.${^{\dag\ddag}}$
\vspace{3.5cm} \normalsize 
               Cedric E. Ginestet${^{ab}}$, \\
               Nicky G. Best${^{c}}$, 
               and Sylvia Richardson${^{c}}$
\end{center}
\begin{center}
\vspace{1cm} 
  \rm ${^a}$ King's College London, Institute of Psychiatry,
  Department of Neuroimaging \\
  \rm $^b$ National Institute of Health Research (NIHR) Biomedical
  Research Centre for Mental Health\\
  \rm $^c$ Department of Epidemiology and Biostatistics, Imperial
  College London\\
\end{center}
\vspace{8cm}
Correspondence concerning this article should be sent to Cedric
Ginestet at the Centre for Neuroimaging Sciences, NIHR Biomedical Research Centre,
Institute of Psychiatry, Box P089, King's College London, 
De Crespigny Park, London, SE5 8AF, UK. Email may be sent to 
\rm cedric.ginestet@kcl.ac.uk
\pagebreak

%-----------------------------------------------
% Spacing.
\onehalfspacing
%\doublespacing
%\setstretch{3} 
\begin{abstract}
   Parameter ensembles or sets of point estimates constitute one of the
   cornerstones of modern statistical practice. This is especially the
   case in Bayesian hierarchical models, where different decision-theoretic frameworks
   can be deployed to summarize such parameter ensembles. 
   The estimation of these parameter ensembles may thus substantially vary
   depending on which inferential goals are prioritised by the
   modeller. In this note, we consider the problem of classifying the
   elements of a parameter ensemble above or below a given threshold. 
   Two threshold classification losses (TCLs) --weighted and
   unweighted-- are formulated. The weighted TCL can be used
   to emphasize the estimation of false positives over
   false negatives or the converse. We prove that the weighted and unweighted TCLs are
   optimized by the ensembles of unit-specific posterior quantiles and posterior
   medians, respectively. In addition, we relate these classification
   loss functions on parameter ensembles to the concepts of posterior
   sensitivity and specificity. Finally, we find some relationships
   between the unweighted TCL and the absolute value loss, which
   explain why both functions are minimized by posterior medians.
\end{abstract}

%%%%%%%%%%%%%%%%%%%%%%%%%%%%%%%%%%%%%%%%%%%%%%%%%%
%%%%%%%%%%%%%%%%%%%%%%%%%%%%%%%%%%%%%%%%%%%%%%%%%%
\section{Introduction}\label{sec:clas intro}
The problem of the optimal classification of a
set of data points into several clusters has 
occupied statisticians and applied mathematicians for several decades
\citep[see][for a overview]{Gordon1999}. As is true for all statistical methods, a
classification is, above all, a summary of the data at hand. When
clustering, the statistician is searching for an optimal
partition of the parameter space into a --generally, known or pre-specified-- number of
classes. The essential ingredient underlying all classifications 
is the minimization of some distance
function, which generally takes the form of a similarity or dissimilarity metric
\citep{Gordon1999}. Optimal classification will then result in a
trade-off between the level of similarity of the within-cluster elements and
the level of dissimilarity of the between-cluster elements. 
In a decision-theoretic framework, such distance functions naturally arise
through the specification of a loss function for the problem at hand. 
The task of computing the optimal partition of the parameter space then
becomes a matter of minimizing the chosen loss function. 

In spatial epidemiology, the issue of classifying areas according to
their levels of risk has been previously investigated by
\citet{Richardson2004}. These authors have shown that areas can be
classified according to the joint posterior distribution of the parameter
ensemble of interest. In particular, a taxonomy can be created by
selecting a decision rule $D(\alpha,C_{\alpha})$ for that purpose, where
$C_{\alpha}$ is a particular threshold, above and below which we
classify the areas in the region of interest. The parameter $\alpha$,
in this decision rule, is the cut-off point associated with $C_{\alpha}$,
which determines the amount of probability mass necessary for an area
to be allocated to the above-threshold category. Thus, an area $i$
with level of risk denoted by $\theta_{i}$ 
will be assigned above the threshold $C_{\alpha}$ if $\p[\theta_{i}>
C_{\alpha}|\by]>\alpha$. \citet{Richardson2004} have therefore
provided a general framework for the classification of areas,
according to their levels of risk. However, this approach is not
satisfactory because it relies on the choice of two co-dependent values
$C_{\alpha}$ and $\alpha$, which can only be selected in an arbitrary
fashion. 

Our perspective in this paper follows the framework adopted by
\citet{Lin2006}, who introduced several loss functions for the identification 
of the elements of a parameter ensemble that represent the proportion of
elements with the highest level of risk. 
Such a classification is based on a particular rank percentile
cut-off denoted $\ga\in [0,1]$, which determines a group of areas of
high-risk. That is, \citet{Lin2006} identified the areas whose percentile
rank is above the cut-off point $\ga$. Our approach, in this paper, is
substantially different since the classification is based on a
real-valued threshold as opposed to a particular rank percentile. In
order to emphasize this distinction, we will refer to our proposed family
of loss functions as threshold classification losses (TCLs).

%%%%%%%%%%%%%%%%%%%%%%%%%%%%%%%%%%%%%%%%%%%%%%%%%%
\section{Classification of Elements in a Parameter Ensemble}\label{sec:class}
We formulate our classification problem within the context of 
Bayesian hierarchical models (BHMs). In its most basic formulation, a
BHM is composed of the following two layers of random variables,
\begin{equation}
      y_{i}\stack{\ind}{\sim} p(y_{i}|\theta_{i},\bsig_{i}), 
      \qq 
      g(\bth) \sim p(\bth|\bxi), 
     \label{eq:bhm}
\end{equation}
for $i=1,\ldots,n$ and where $g(\cdot)$ is a transformation of $\bth$,
which may be defined as a link function as commonly used
in generalised linear models \citep[see][]{McCullagh1989}. The vector
of real-valued parameters, $\bth:=\lb\theta_{1},\ldots,\theta_{n}\rb$,
will be referred to as a \ti{parameter ensemble}.

%%%%%%%%%%%%%%%%%%%%%%%%%%%%%%%%%%%%%%%%%%%%%%%%%%
\subsection{Threshold Classification Loss}\label{sec:tcl}
For some cut-off point $C\in\R$, we define the penalties associated with
the two different types of misclassification. Following standard statistical terminology, we
will express such misclassifications in terms of false positives (FPs)
and false negatives (FNs). These concepts are formally described as
\begin{equation}\label{eq:abba1}
    \op{FP}(C,\theta,\theta^{\op{est}}) := \cI\lt\lb \theta \leq C, \theta^{\op{est}} > C
    \rt\rb,
     \q\te{and}\q
    \op{FN}(C,\theta,\theta^{\op{est}}) :=  \cI\lt\lb  \theta > C, \theta^{\op{est}} \leq C \rt\rb,
\end{equation}
where $\theta$ represents the parameter of interest and $\theta^{\op{est}}$ is a
candidate estimate. This corresponds to the occurrence of a false positive
(type I error) and a false negative (type II error), respectively. 

For the decision problem to be fully specified, we need to choose
a loss function based on the sets of unit-specific FPs and FNs. The
$p$-weighted threshold classification loss ($\op{TCL}_{p}$) function
is then defined as
\begin{equation}
     \op{TCL}_{p}(C,\bth,\bm\theta^{\op{est}}) := \frac{1}{n} \sum_{i=1}^{n}
     p\op{FP}(C,\theta_{i},\theta^{\op{est}}_{i}) + (1-p)\op{FN}(C,\theta_{i},\theta^{\op{est}}_{i}).     
     \label{eq:ptcl}     
\end{equation}
One of the advantages of the choice of $\op{TCL}_{p}$ for quantifying
the misclassifications of the elements of a parameter ensemble is that
it is normalised, in the sense that $\op{TCL}_{p}(C,\bth,\bm\theta^{\op{est}})\in [0,1]$
for any choice of $C$ and $p$. Our main result in this paper is
the following minimization. 
%%%%%%%%%%%%%%%%%%%%%%%%%%%%%%%%%%%%%%%%%%%%%%%%%%
\begin{pro}\label{pro:tcl}
  For some parameter ensemble $\bth$, and given a real-valued
  threshold $C\in\R$ and $p\in [0,1]$, we have the following optimal estimator under
  weighted TCL, 
  \begin{equation}
    \bth^{\op{TCL}}_{(1-p)} = \argmin_{\bm\theta^{\op{est}}}
    \E\lt[\op{TCL}_{p}(C,\bth,\bm\theta^{\op{est}})|\by\rt],
    \label{eq:tcl minimizer}
  \end{equation}
  where $\bth^{\op{TCL}}_{(1-p)}$ is the vector of posterior $(1-p)$-quantiles defined
  as 
  \begin{equation}
      \bth^{\op{TCL}}_{(1-p)}:=\lt\lb Q_{\theta_{1}|\by}(1-p),\ldots,Q_{\theta_{n}|\by}(1-p)\rt\rb,
  \end{equation}
  where $Q_{\theta_{i}|\by}(1-p)$ denotes the posterior $(1-p)$-quantile of
  the $i\tth$ element, $\theta_{i}$, in the parameter ensemble. Moreover,
  $\bth^{\op{TCL}}_{(1-p)}$ is not unique. 
\end{pro}
%%%%%%%%%%%%%%%%%%%%%%%%%%%%%%%%%%%%%%%%%%%%%%%%%%
We prove this result by exhaustion in three cases.
The full proof is reported in Appendix A. 
Note that the fact that $\op{TCL}_{p}$ is minimized
by $\bth^{\op{TCL}}_{(1-p)}$ and not $\bth^{\op{TCL}}_{(p)}$ is solely
a consequence of our choice of definition for the $\op{TCL}_{p}$
function. If the weighting of the FPs and FNs had been $(1-p)$ and
$p$, respectively, then the optimal minimizer of that function would
indeed be a vector of posterior $p$-quantiles. 

%%%%%%%%%%%%%%%%%%%%%%%%%%%%%%%%%%%%%%%%%%%%%%%%%%
\subsection{Unweighted Threshold Classification Loss}\label{sec:unweighted tcl}
We now specialize this result to the unweighted TCL family, which is
defined analogously to equation (\ref{eq:ptcl}), as follows,
\begin{equation}
     \op{TCL}(C,\bth,\bm\theta^{\op{est}}) := \frac{1}{n} \sum_{i=1}^{n}
     \op{FP}(C,\theta_{i},\theta^{\op{est}}_{i}) + \op{FN}(C,\theta_{i},\theta^{\op{est}}_{i}).
     \label{eq:tcl}
\end{equation}
The minimizer of this loss function can be shown to be trivially
equivalent to the minimizer of $\op{TCL}_{0.5}$. That is, we have
\begin{equation}
      \argmin_{\bm\theta^{\op{est}}} \E[\op{TCL}(C,\bth,\bm\theta^{\op{est}})|\by] =
      \argmin_{\bm\theta^{\op{est}}} \E[\op{TCL}_{0.5}(C,\bth,\bm\theta^{\op{est}})|\by],
\end{equation}
for every $C$, which therefore proves the following corollary.
\begin{cor}\label{cor:tcl}
  For some parameter ensemble $\bth$ and $C\in\R$, the minimizer of
  the posterior expected TCL is 
  \begin{equation}
       \bth^{\op{med}} := \bth^{\op{TCL}}_{(0.5)} =\lt\lb
        Q_{\theta_{1}|\by}(0.5),\ldots,Q_{\theta_{n}|\by}(0.5)\rt\rb, 
  \end{equation}
  and this optimal estimator is not unique. 
\end{cor}
The posterior expected loss under the unweighted TCL function takes the following
form,
\begin{equation}
    \E\lt[\op{TCL}(C,\bth,\bm\theta^{\op{est}})|\by\rt]
    = \frac{1}{n}\sum_{i=1}^{n} \int\limits_{-\infty}^{C} d\p[\theta_{i}|\by]
    \cI\lt\lb \theta^{\op{est}}_{i} > C\rt\rb
    + \int\limits_{C}^{+\infty}d\p[\theta_{i}|\by]
    \cI\lt\lb \theta^{\op{est}}_{i} \leq C\rt\rb,
    \label{eq:posterior tcl}
\end{equation}
whose formulae is derived using $\cI\lt\lb \theta \leq C,\theta^{\op{est}} > C\rt\rb = \cI\lt\lb \theta \leq
C\rt\rb\cI\lt\lb\theta^{\op{est}} > C\rt\rb$. It is of special importance to note
that when using the posterior TCL, any classification --correct or
incorrect-- will incur a penalty. The \ti{size} of that penalty,
however, varies substantially depending on whether or not the
classification is correct. A true positive can be distinguished from a false positive, 
by the fact that the former will only incur a small penalty
proportional to the posterior probability of the parameter to be
below the chosen cut-off point $C$. 

%%%%%%%%%%%%%%%%%%%%%%%%%%%%%%%%%%%%%%%%%%%%%%%%%
%%%%%%%%%%%%%%%%%%%%%%%%%%%%%%%%%%%%%%%%%%%%%%%%%%
\subsection{Relationship with Posterior Sensitivity and Specificity}\label{sec:sense and sensitivity}
Our chosen decision-theoretic framework for classification has the added benefit of being
readily comparable to conventional measures of
classification errors widely used in the context of test theory.
For our purpose, we will define the Bayesian sensitivity 
of a classification estimator $\bm\theta^{\op{est}}$, also
referred to as the posterior true positive rate (TPR), as follows
\begin{equation}
    \op{TPR}(C,\bth,\bm\theta^{\op{est}})
    := \frac{\sum_{i=1}^{n} \E[\op{TP}(C,\theta_{i},\theta^{\op{est}}_{i})|\by]}
    {\sum_{i=1}^{n} \p[\theta_{i} > C |\by]},
    \label{eq:tpr}
\end{equation}
where the expectations are taken with respect to the joint posterior
distribution of $\bth$. Similarly, the Bayesian specificity, or
posterior true negative rate (TNR), will be defined as 
\begin{equation}
    \op{TNR}(C,\bth,\bm\theta^{\op{est}})
    := \frac{\sum_{i=1}^{n}\E[\op{TN}(C,\theta_{i},\theta^{\op{est}}_{i})|\by]}
    {\sum_{i=1}^{n} \p[\theta_{i} \leq C |\by]},
    \label{eq:tnr}
\end{equation}
where in both definitions, we have used $\op{TP}(C,\theta_{i},\theta^{\op{est}}_{i}) := 
\cI\lt\lb \theta_{i} > C , \theta^{\op{est}}_{i} > C \rt\rb$
and $\op{TN}(C,\theta_{i},\theta^{\op{est}}_{i}) := \cI\lt\lb  \theta_{i} \leq C,
\theta^{\op{est}}_{i} \leq C\rt\rb$. It then follows that we can formulate the
relationship between the posterior expected TCL and the Bayesian
sensitivity and specificity as
\begin{equation}
  \E[\op{TCL}(C,\bth,\bm\theta^{\op{est}})|\by]     
    = \frac{1}{n}\op{FPR}(C,\bth,\bm\theta^{\op{est}})\sum_{i=1}^{n} \p[\theta_{i} \leq C |\by]
    + \frac{1}{n}\op{FNR}(C,\bth,\bm\theta^{\op{est}})\sum_{i=1}^{n} \p[\theta_{i} > C |\by]. 
    \notag
\end{equation}
where $\op{FPR}(C,\bth,\bm\theta^{\op{est}}) := 1- \op{TNR}(C,\bth,\bm\theta^{\op{est}})$
and $\op{FNR}(C,\bth,\bm\theta^{\op{est}}) := 1- \op{TPR}(C,\bth,\bm\theta^{\op{est}})$.

%%%%%%%%%%%%%%%%%%%%%%%%%%%%%%%%%%%%%%%%%%%%%%%%%
%%%%%%%%%%%%%%%%%%%%%%%%%%%%%%%%%%%%%%%%%%%%%%%%%%
\section{Conclusion}\label{sec:conclusion}
The fact that the posterior median is the minimizer of the 
posterior expected absolute value loss (AVL) function is well-known
\citet{Berger1980}. That is, the posterior median minimizes the
posterior expected AVL, where $\op{AVL}(\theta,\theta^{\op{est}}):=|\theta -
\theta^{\op{est}}|$. One may therefore ask whether there is link
between the minimization of the AVL function, which is an estimation loss
and the classification loss function described in this paper. 
The proof of the optimality of the posterior median under AVL proceeds by considering 
whether $\theta^{\op{med}}-\theta^{\op{est}}\gtreqless 0$. This leads
to a proof by exhaustion in three cases, which includes the trivial case 
where $\theta^{\op{med}}$ and $\theta^{\op{est}}$ are equal.
Similarly, in the proof of proposition \ref{pro:tcl}, we have also obtained three cases,
which are based on the relationships between the $\theta_{i}$'s and
$\theta_{i}^{\op{est}}$'s with respect to $C$. However, note that by
subtracting $\theta_{i}^{(1-p)} \leq C$ from
$C<\theta_{i}^{\op{est}}$ and ignoring null sets, we obtain 
$\theta_{i}^{(1-p)} - \theta_{i}^{\op{est}} < 0$,
for the second case. Similarly, a subtraction of the hypotheses of the
third case gives $\theta_{i}^{(1-p)} - \theta_{i}^{\op{est}} > 0$
for the third case, which therefore highlights the relationship
between the optimization of the AVL and the TCL functions. 

%%%%%%%%%%%%%%%%%%%%%%%%%%%%%%%%%%%%%%%%%%%%%%%%%
%%%%%%%%%%%%%%%%%%%%%%%%%%%%%%%%%%%%%%%%%%%%%%%%%
\setcounter{secnumdepth}{-2}
\pagebreak
{\singlespacing
\section{Appendix A: Proof of TCL Minimization}
%%%%%%%%%%%%%%%%%%%%%%%%%%%%%%%%%%%%%%%%%%%%%%%%%%
\sub{Proof of proposition \ref{pro:tcl} on page \pageref{pro:tcl}.}
Let $\rho_{p}(C,\bth,\bth^{\op{est}})$ denote $\E[\op{TCL}_{p}(C,\bth,\bth^{\op{est}})|\by]$.
We prove the result by exhaustion over three cases. In order to prove that 
\begin{equation}
   \rho_{p}(C,\bth,\bth^{(1-p)})\leq\rho_{p}(C,\bth,\bth^{\op{est}}), 
\end{equation}
for any $\bth^{\op{est}}\in\bTh$ with $\theta_{i}^{(1-p)}:=Q_{\theta_{i}|\by}(1-p)$, it suffices to show that 
$\rho_{p}(C,\theta_{i},\theta_{i}^{(1-p)})\leq\rho_{p}(C,\theta_{i},\theta_{i}^{\op{est}})$
holds, for every $i=1,\ldots,n$. Expanding these unit-specific risks,
\begin{equation}
  \begin{aligned}
    p\cI\lb&\theta_{i}^{(1-p)}>C\rb\p\lt[\theta_{i}\leq
    C|\by\rt] + (1-p)\cI\lb\theta_{i}^{(1-p)}\leq
    C\rb\p\lt[\theta_{i}> C|\by\rt] \\
    &\leq\,
    p\cI\lb\theta_{i}^{\op{est}}>C\rb\p\lt[\theta_{i}\leq
    C|\by\rt] + (1-p)\cI\lb\theta_{i}^{\op{est}}\leq
    C\rb\p\lt[\theta_{i}> C|\by\rt].
   \label{eq:expansion}
  \end{aligned}
\end{equation}

Now, fix $C$ and $p\in[0,1]$ to arbitrary values. Then, for any
point estimate $\theta_{i}^{\op{est}}$, we have
\begin{equation}
  \rho_{p}(C,\theta_{i},\theta_{i}^{\op{est}}) = 
  \begin{cases}
    p\p[\theta_{i}\leq C|\by], & \te{if }\theta_{i}^{\op{est}}> C,\\
    (1-p)\p[\theta_{i}> C|\by], & \te{if }\theta_{i}^{\op{est}}\leq C.
  \end{cases}
\end{equation}
The optimality of $\theta^{(1-p)}_{i}$ over
$\theta^{\op{est}}_{i}$ as a point estimate is therefore directly
dependent on the relationships between $\theta_{i}^{(1-p)}$ and $C$,
and between $\theta_{i}^{\op{est}}$ and $C$. This determines the
following three cases:
\begin{description}
  \item[i.] If $\theta_{i}^{(1-p)}$ and $\theta_{i}^{\op{est}}$ are
    on the same side of $C$, then clearly, 
    \begin{equation}
    \rho_{p}(C,\theta_{i},\theta_{i}^{(1-p)}) = \rho_{p}(C,\theta_{i},\theta_{i}^{\op{est}}),
    \label{eq:case1}
    \end{equation}
  \item[ii.] If $\theta_{i}^{(1-p)} \leq C$ and $\theta_{i}^{\op{est}}> C$, then,
    \begin{equation}
    \rho_{p}(C,\theta_{i},\theta_{i}^{(1-p)}) 
    = (1-p)\p[\theta_{i}> C|\by]
    \,\leq\,
    p\p[\theta_{i}\leq C|\by] =
    \rho_{p}(C,\theta_{i},\theta_{i}^{\op{est}}),
    \label{eq:case3}
    \end{equation}
  \item[iii.] If $\theta_{i}^{(1-p)} > C$ and $\theta_{i}^{\op{est}}\leq C$, then, 
    \begin{equation}
    \rho_{p}(C,\theta_{i},\theta_{i}^{(1-p)}) 
    = p\p[\theta_{i}\leq C|\by]
    \,<\,
    (1-p)\p[\theta_{i}> C|\by] =
    \rho_{p}(C,\theta_{i},\theta_{i}^{\op{est}}),     
    \label{eq:case4}
    \end{equation} 
\end{description}
Equation (\ref{eq:case1}) follows directly from
an application of the result in (\ref{eq:expansion}), and cases two
and three follow from consideration of the following
relationship:
\begin{equation}
  p\p[\theta_{i}\leq C|\by] \gtreqless (1-p)\p[\theta_{i}>
  C|\by],
  \label{eq:grteqless1}
\end{equation}
where $\gtreqless$ means either $<$, $=$ or $>$. 
Using $\p[\theta_{i}> C|\by]=1-\p[\theta_{i}\leq C|\by]$, this gives
\begin{equation}
    \p[\theta_{i}\leq C|\by] = F_{\theta_{i}|\by}(C) \gtreqless 1-p.
     \label{eq:grteqless3}
\end{equation}
Here, $F_{\theta_{i}|\by}$ is the posterior CDF
of $\theta_{i}$. Therefore, we have
\begin{equation}
     C \gtreqless F^{-1}_{\theta_{i}|\by}(1-p) =:
     Q_{\theta_{i}|\by}(1-p) :=: \theta_{i}^{(1-p)},
     \label{eq:grteqless2}
\end{equation}
where $\gtreqless$ takes the same value in equations
(\ref{eq:grteqless1}), (\ref{eq:grteqless3}) and (\ref{eq:grteqless2}).

This proves the optimality of $\bth^{(1-p)}$. Moreover, since one can
construct a vector of point estimates $\theta_{i}^{\op{est}}$
satisfying $\theta_{i}^{\op{est}}\gtreqless C$, whenever 
$\theta^{(1-p)}_{i}\gtreqless C$, for every $i$, it
then follows that $\bth^{(1-p)}$ is not unique.
}

%%%%%%%%%%%%%%%%%%%%%%%%%%%%%%%%%%%%%%%%%%%%%%%%%
%%%%%%%%%%%%%%%%%%%%%%%%%%%%%%%%%%%%%%%%%%%%%%%%%%
%%%%%%%%%%%%%%%%%%%%%%%%%%%%%%%%%%%%%%%%%%%%%%%%%%
\section{Acknowledgments} 
This work was supported by a fellowship from the UK National Institute
for Health Research (NIHR) Biomedical Research Centre for Mental
Health (BRC-MH) at the South London and Maudsley NHS Foundation Trust
and King's College London. This work has also been funded by
the Guy's and St Thomas' Charitable Foundation as well as the South
London and Maudsley Trustees. 
% We would also like to thank two anonymous reviewers for their valuable inputs.

% references --------------------------------------------------
\small
\singlespacing
\addcontentsline{toc}{section}{References}
\bibliography{/home/cgineste/ref/bibtex/Statistics,%
             /home/cgineste/ref/bibtex/Neuroscience}
\bibliographystyle{oupced3}

% Index -------------------------------------------------------

\addcontentsline{toc}{section}{Index}
%\printindex

\end{document}